\documentclass[twocolumn,aps,pra,showpacs]{revtex4}
\usepackage[T1]{fontenc}
\usepackage{mathrsfs}
\usepackage{bm}
\usepackage{amsmath}
\usepackage{amssymb}
\usepackage{graphicx}
\usepackage{esint}

\makeatletter
\@ifundefined{textcolor}{}
{%
 \definecolor{BLACK}{gray}{0}
 \definecolor{WHITE}{gray}{1}
 \definecolor{RED}{rgb}{1,0,0}
 \definecolor{GREEN}{rgb}{0,1,0}
 \definecolor{BLUE}{rgb}{0,0,1}
 \definecolor{CYAN}{cmyk}{1,0,0,0}
 \definecolor{MAGENTA}{cmyk}{0,1,0,0}
 \definecolor{YELLOW}{cmyk}{0,0,1,0}
}

\makeatother

\usepackage[english]{babel}
\begin{document}

\title{Superfluid density and Berezinskii-Kosterlitz-Thouless transition
of a spin-orbit coupled Fulde-Ferrell superfluid}

\author{Ye Cao$^{1,2}$}

\author{Xia-Ji Liu$^{1}$}

\author{Lianyi He$^{3}$}

\author{Gui-Lu Long$^{2,4,5}$}

\author{Hui Hu$^{1}$}

\email{hhu@swin.edu.au}

\affiliation{$^{1}$Centre for Quantum and Optical Science, Swinburne University
of Technology, Melbourne 3122, Australia}

\affiliation{$^{2}$State Key Laboratory of Low-dimensional Quantum Physics and
Department of Physics, Tsinghua University, Beijing 100084, P. R.
China}

\affiliation{$^{3}$Theoretical Division, Los Alamos National Laboratory, Los
Alamos, NM87545, USA}

\affiliation{$^{4}$Collaborative Innovation Center of Quantum Matter, Beijing
100084, P. R. China}

\affiliation{$^{5}$Tsinghua National Laboratory for Information Science and Technology,
Beijing 100084, P. R. China}

\date{\today}
\begin{abstract}
We theoretically investigate the superfluid density and Berezinskii-Kosterlitz-Thouless
(BKT) transition of a two-dimensional Rashba spin-orbit coupled atomic
Fermi gas with both in-plane and out-of-plane Zeeman fields. It was
recently predicted that, by tuning the two Zeeman fields, the system
may exhibit different exotic Fulde-Ferrell (FF) superfluid phases,
including the gapped FF, gapless FF, gapless topological FF and gapped
topological FF states. Due to the FF paring, we show that the superfluid
density (tensor) of the system becomes anisotropic. When an in-plane
Zeeman field is applied along the \textit{x}-direction, the tensor
component along the \textit{y}-direction $n_{s,yy}$ is generally
larger than $n_{s,xx}$ in most parameter space. At zero temperature,
there is always a discontinuity jump in $n_{s,xx}$ as the system
evolves from a gapped FF into a gapless FF state. With increasing
temperature, such a jump is gradually washed out. The critical BKT
temperature has been calculated as functions of the spin-orbit coupling
strength, interatomic interaction strength, in-plane and out-of-plane
Zeeman fields. We predict that the novel FF superfluid phases have
a significant critical BKT temperature, typically at the order of
$0.1T_{F}$, where $T_{F}$ is the Fermi degenerate temperature. Therefore,
their observation is within the reach of current experimental techniques
in cold-atom laboratories. 
\end{abstract}

\pacs{05.30.Fk, 03.75.Hh, 03.75.Ss, 67.85.-d}

\maketitle

\section{Introduction}

Over the past decade, the technique of manipulating ultracold atomic
Fermi gases has been well developed and it offers a physical reality
to pursue an exotic pairing mechanism, which is referred to as Fulde-Ferrell-Larkin-Ovchinnikov
(FFLO) states \cite{Fulde1964,Larkin1964} and has attracted impressive
attentions in different physical areas \cite{Casalbuoni2004,Uji2006,Kenzelmann2008,Liao2010,Gerber2014}.
In spin-imbalanced Fermi gases, the standard Bardeen-Cooper-Schrieffer
(BCS) pairing is not favorable against the FFLO pairing with a finite
center-of-mass momentum. Although there is no unambiguous experimental
conclusion for the FFLO superfluidity, strong evidence has been seen
in a Fermi cloud of $^{6}$Li atoms confined in quasi-one-dimensional
harmonic traps near a crossover from a Bose-Einstein condensate (BEC)
to a BCS superfluid \cite{Liao2010,Orso2007,Hu2007,Liu2007,Liu2008,Guan2007}. 

The FFLO pairing is also favored by spin-orbit coupling \cite{YipReview,Barzykin2002,Agterberg2007,Dimitrova2007,Michaeli2012}.
Motivated by the recent experimental realization of a synthetic spin-orbit
coupling with equal weight combination of Rashba and Dresselhaus components
\cite{Lin2011,Wang2012,Cheuk2012,Fu2013}, FF superfluidity - a specific
form of the FFLO superfluidity - has been theoretically investigated
in spin-orbit coupled atomic Fermi gases \cite{reviewSOC,Dong2013,Zheng2013,Wu2013,Liu2013a,Dong2013NJP,Zhou2013,Iskin2013,Shenoy2013}.
In the case of a Rasbha spin-orbit coupling, topological superfluidity
is argued to be achievable \cite{Zhang2008,Sato2009,Sau2010,Oreg2010,Zhu2011,Liu2012a,Liu2012b,Wei2012},
although the underlying pairing is of \textit{s}-wave character. It
turns out that the topological superfluidity and FF superfluidity
are compatible. As a result, novel topological FF superfluids have
also been proposed \cite{Chen2013,Liu2013b,Qu2013,Zhang2013,Cao2014,Hu2014,Jiang2014}.
In particular, in a recent Letter, some of us have predicted that
a \emph{gapless} topological FF superfluid may appear in a two-dimensional
(2D) spin-orbit coupled atomic Fermi gas with both in-plane and out-of-plane
Zeeman fields \cite{Cao2014}. The purpose of the present work is
to provide more details about such an interesting superfluid phase
and to discuss its thermodynamic stability by considering the superfluid
density and superfluid transition temperature.

It is well known that at finite temperatures the superfluidity of
2D atomic Fermi gases is characterized by the vortex-antivortex (V-AV)
binding. The relevant mechanism is the Berezinskii-Kosterlitz-Thouless
(BKT) transition occurring at a characteristic temperature $T_{\mathrm{{BKT}}}$
\cite{Berezinskii1971,Kosterlitz1972}. Below the critical BKT temperature,
a V-AV binding state has a lower free energy and hence superfluidity
emerges. The BKT transition was theoretically investigated long time
ago in a 2D fermionic system without spin-orbit coupling \cite{Randeria1989,Gusynin1999,Loktev2001}.
Following the recent experimental advances, there have been several
theoretical investigations about the superfluid density and critical
BKT temperature in 2D spin-orbit coupled Fermi gases with BCS pairing
\cite{He2012,Gong2012,He2013}. In the case of a large out-of-plane
Zeeman field, the temperature region for experimentally observing
topological BCS superfluids and related Majorana fermions has been
discussed \cite{Gong2012,He2013}. However, the BKT physics of a spin-orbit
coupled FF superfluid - which can be either gapped or gapless, topologically
trivial or non-trivial - has so far not been addressed.

In this work, we explore this interesting issue and study the superfluid
density tensor and BKT transition of a 2D Rasbha spin-orbit coupled
Fermi gas in the presence of both in-plane and out-of-plane Zeeman
fields. By calculating the superfluid density tensor, we obtain the
superfluid phase stiffness as functions of the temperature, spin-orbit
coupling strength, binding energy (that characterizes the interatomic
interaction strength), in-plane and out-of-plane Zeeman fields. This
allows us to determine the critical BKT temperature of the system
in four different FF superfluid phases \cite{Cao2014}, with a given
set of parameters. 

Our main results may be summarized as follows: (i) At zero temperature
with an applied in-plane Zeeman field in the \textit{x}-direction,
the component $n_{s,xx}$ of the superfluid density tensor always
changes discontinuously when the system continuously evolves from
a gapped FF into a gapless FF phase. The component $n_{s,yy}$ is
larger than $n_{s,xx}$ except for a narrow parameter space where
the FF momentum is sufficiently large. The two components of the superfluid
density tensor decrease monotonically as the temperature increases.
(ii) All the four FF superfluid phases have significant critical BKT
temperature, except for the parameter region with very small spin-orbit
coupling and/or binding energy, or with very large in-plane and/or
out-of-plane Zeeman fields. The critical BKT temperature can be enhanced
by increasing the binding energy. But it does not increase monotonically
as the spin-orbit coupling strength increases.

The rest of the paper is organized as follows. In the next section,
we briefly describe the mean-field theoretical framework, and clarify
the BKT physics in two dimensions and the related Kosterlitz-Thouless-Nelson
(KT-Nelson) criterion for phase transition. Then, we present the expressions
for the superfluid density tensor and superfluid phase stiffness.
The critical BKT temperature is determined by applying the KT-Nelson
criterion. In Sec. III, we first present the finite-temperature phase
diagram of the system and then discuss in detail the results on the
superfluid density tensor and critical BKT temperature. Finally, Sec.
IV is devoted to the conclusions and outlooks.

\section{Model Hamiltonian and mean-field theory}

We start by considering a 2D spin-orbit coupled two-component Fermi
gas near a broad Feshbach resonance with the Rashba spin-orbit coupling
$\lambda\bm{\hat{\sigma}}\cdot\bm{\hat{\mathrm{k}}}$, in-plane ($h_{x}$)
and out-of-plane ($h_{z}$) Zeeman fields \cite{notation}. The system
can be well described by the following single-channel Hamiltonian,
\begin{equation}
\text{\ensuremath{\mathcal{H}}}=\int d{\bf r}\left[\mathcal{H}_{0}+\mathcal{H}_{int}\right],\label{eq:totHami}
\end{equation}
where

\begin{equation}
{\cal H}_{0}=\psi^{\dagger}(\bm{r})\left(\hat{\xi}_{{\bf k}}+\lambda\bm{\hat{\sigma}}\cdot\bm{\hat{\mathrm{k}}}-h_{z}\hat{\sigma}_{z}-h_{x}\hat{\sigma}_{x}\right)\psi(\bm{r})\label{eq:spHami}
\end{equation}
is the single-particle Hamiltonian and 
\begin{equation}
\text{\ensuremath{\mathcal{H}}}_{int}=U_{0}\psi_{\uparrow}^{\dagger}({\bf r})\psi_{\downarrow}^{\dagger}({\bf r})\psi_{\downarrow}({\bf r})\psi_{\uparrow}({\bf r})
\end{equation}
is the density of interaction Hamiltonian in which the bare interaction
strength $U_{0}$ is to be regularized as 
\begin{equation}
\frac{1}{U_{0}}=-\frac{1}{\mathcal{S}}\sum_{\mathbf{k}}\frac{1}{\hbar^{2}\mathbf{k}^{2}/m+E_{b}},
\end{equation}
with $\mathcal{S}$ being the area of the system and $E_{b}$ the
two-particle binding energy that physically characterizes the interaction
strength. In the single-particle Hamiltonian, $\lambda$ is the Rashba
spin-orbit coupling strength and we have used the following notations:
(1) $\hat{\xi}_{{\bf k}}\equiv-\hbar^{2}\nabla^{2}/(2m)-\mu$ with
the atomic mass $m$ and chemical potential $\mu$; (2) $\bm{\mathrm{\hat{k}}}=(\hat{k}_{x},\hat{k}_{y})$,
where $\hat{k}_{x}=-i\partial_{x}$ and $\hat{k}_{y}=-i\partial_{y}$
are momentum operators; and (3) $\bm{\hat{\sigma}}=(\hat{\sigma}_{x},\hat{\sigma}_{y})$,
the Pauli matrices. We have also used $\psi({\bf r})=[\psi_{\uparrow}(\bm{r}),\psi_{\downarrow}(\bm{r})]^{T}$
($\psi^{\dagger}(\bm{r})=[\psi_{\uparrow}^{\dagger}(\bm{r}),\psi_{\downarrow}^{\dagger}(\bm{r})]$)
to collectively denote the fermion field operator for creating (annihilating)
an atom at ${\bf r}$ with a specific spin $\sigma=\uparrow,\downarrow$.

\subsection{Mean-field theory}

We solve the model Hamiltonian Eq. (\ref{eq:totHami}) by using the
functional path-integral approach \cite{reviewSOC,He2013,Hu2011,Jiang2011}.
At the inverse finite temperature $\beta=1/(k_{B}T)$, the partition
function can be written as,

\begin{equation}
\text{\ensuremath{\mathcal{Z}}}=\int\mathcal{\mathcal{D}}\psi\left(\mathbf{r},\tau\right)\mathcal{D}\bar{\psi}\left(\mathbf{r},\tau\right)\exp\left\{ -\mathcal{A\left[\psi,\bar{\psi}\right]}\right\} ,\label{eq:partition1}
\end{equation}
where

\begin{equation}
\mathcal{A}\left[\psi,\bar{\psi}\right]=\int_{0}^{\beta}d\tau\int d\bm{r}\bar{\psi}\partial_{\tau}\psi+\int_{0}^{\beta}d\tau\mathcal{H}\left(\psi,\bar{\psi}\right).\label{eq:action}
\end{equation}
Here, the field operators $\psi$ and $\psi^{\dagger}$ in the model
Hamiltonian $\mathcal{H}$ have been replaced with the corresponding
Grassmann variables $\psi(\mathbf{r},\tau)$ and $\bar{\psi}(\mathbf{r},\tau)$,
respectively. Following the standard procedure \cite{reviewSOC},
the interaction term in the Hamiltonian is decoupled using the Hubbard-Stratonovich
transformation. Introducing the auxiliary complex pairing field $\phi(\mathbf{r},\tau)=-U_{0}\psi_{\downarrow}(\mathbf{r},\tau)\psi_{\uparrow}(\mathbf{r},\tau)$,
and integrating out the Grassmann fields, the partition function becomes

\begin{equation}
\text{\ensuremath{\mathcal{Z}}}=\int\mathcal{\mathcal{D}\phi}\left(\mathbf{r},\tau\right)\mathcal{D}\bar{\phi}\left(\mathbf{r},\tau\right)\exp\left\{ -\mathcal{A}_{eff}\left[\phi,\bar{\phi}\right]\right\} ,\label{eq:partition2}
\end{equation}
where in the saddle-point approximation (i.e., mean-field treatment
by replacing $\phi(\mathbf{r},\tau)$ with a static pairing field
$\Delta(\mathbf{r})$), the effective action $\text{\ensuremath{\mathcal{A}}}_{eff}$
takes the form, 

\begin{equation}
\text{\ensuremath{\mathcal{A}}}_{mf}=\beta\sum_{\mathbf{k}}\hat{\xi}_{\mathbf{k}}-\int_{0}^{\beta}d\tau\int d\bm{r}\frac{\left|\Delta\right|{}^{2}}{U_{0}}-\frac{1}{2}\textrm{Tr}\ln\left[-G^{-1}\right].\label{eq:effaction}
\end{equation}
In the above expression, $G^{-1}\left(\mathbf{r},\tau\right)=-\partial_{\tau}-\mathcal{H}_{BdG}$
is the inverse single-particle Green function in the Nambu-Gorkov
representation, with a mean-field Bogoliubov Hamiltonian,
\begin{equation}
\mathcal{H}_{BdG}=\left[\begin{array}{cc}
H_{0}(\mathbf{\hat{k}}) & -i\Delta(\mathbf{r})\hat{\sigma}_{y}\\
i\Delta(\mathbf{r})\hat{\sigma}_{y} & -H_{0}^{*}\left(-\mathbf{\hat{k}}\right)
\end{array}\right],
\end{equation}
where $H_{0}\equiv\hat{\xi}_{{\bf k}}+\lambda\bm{\hat{\sigma}}\cdot\bm{\hat{\mathrm{k}}}-h_{z}\hat{\sigma}_{z}-h_{x}\hat{\sigma}_{x}$.
In the presence of the in-plane Zeeman field $h_{x}$, it is known
that the pairing field takes the FF form $\Delta(\mathbf{r})=\Delta e^{iQx}$,
with a finite center-of-mass momentum of the pairs $\mathbf{Q}=Q\mathbf{e}_{x}$
\cite{Zheng2013,Wu2013,Liu2013a,Dong2013NJP,Zhou2013}. This helical
phase was earlier studied in the context of noncentrosymmetric superconductors
\cite{Agterberg2007,Dimitrova2007}. The resulting mean-field thermodynamic
potential $\Omega_{mf}=k_{B}T\mathcal{A}_{mf}$ reads,

\begin{equation}
\Omega_{mf}=\sum_{\mathbf{k}}\hat{\xi}_{\mathbf{k}}-\mathcal{S}\frac{\Delta^{2}}{U_{0}}-\frac{k_{B}T}{2}\sum_{\mathbf{k},i\omega_{m}}\ln\det\left[-G^{-1}\left(\mathbf{k},i\omega_{m}\right)\right],\label{eq:thermaldynamic}
\end{equation}
where $G^{-1}(\mathbf{k},i\omega_{m})$ is the inverse Green function
in momentum space and $\omega_{m}=\pi(2m+1)/\beta$ with integer $m$
is the fermionic Matsubara frequency. Making use of the inherent particle-hole
symmetry of the BdG Hamiltonian, we find that,
\begin{equation}
\det\left[-G^{-1}\left(\mathbf{k},i\omega_{m}\right)\right]=\prod_{\eta=1,2}\left[\left(i\omega_{n}\right){}^{2}-\left(E_{\bm{\mathrm{k}}\eta}^{\nu=+}\right){}^{2}\right],
\end{equation}
where $E_{\mathrm{\bm{k}\eta}}^{\nu}$ is the quasi-particle energy,
obtained by diagonalizing $\mathcal{H}_{BdG}$ with the FF pairing
field $\Delta(\mathbf{r})=\Delta e^{iQx}$ \cite{Liu2013a,Dong2013NJP}.
The superscript $\nu\in(+,-)$ represents the particle ($+$) or hole
($-$) branch and the subscript $\text{\ensuremath{\eta\in}(1,2)}$
denotes the upper ($1$) or lower ($2$) branch split by the spin-orbit
coupling \cite{Liu2013a,Hu2011,Jiang2011}. By summing over the Matsubara
frequency, the mean-field thermodynamic potential takes the form,

\begin{eqnarray}
\Omega_{mf} & = & \frac{1}{2}\sum_{\mathbf{\bm{k}}}\left(\xi_{\mathbf{\bm{k}}+\mathbf{\bm{Q}}/2}+\xi_{\mathbf{\bm{k}}-\mathbf{\bm{Q}}/2}\right)-\frac{1}{2}\sum_{\mathbf{\bm{k}\eta}}|E_{\bm{\mathrm{\bm{k}}}\eta}^{+}|\nonumber \\
 &  & -k_{B}T\sum_{\mathbf{k\eta}}\ln\left(1+e^{-|E_{\mathbf{k}\eta}^{+}|/k_{B}T}\right)-\mathcal{S}\frac{\Delta^{2}}{U_{0}}.
\end{eqnarray}
Here the term $\sum_{\mathbf{k}}\hat{\xi}_{\mathbf{k}}$ is replaced
by $(1/2)\sum_{\mathbf{k}}(\xi_{\mathbf{k}+\mathbf{Q}/2}+\xi_{\mathbf{k}-\mathbf{Q}/2})$,
in order to cancel the leading divergence of the term $(1/2)\sum_{\mathbf{\bm{k}\eta}}|E_{\bm{\mathrm{\bm{k}}}\eta}^{+}|$.

For a given set of parameters, for example, the temperature $T$,
binding energy $E_{b}$ etc., different superfluid phases can be determined
using the self-consistent stationary conditions: 
\begin{eqnarray}
\frac{\partial\Omega_{mf}}{\partial\Delta} & = & 0,\\
\frac{\partial\Omega_{mf}}{\partial Q} & = & 0,
\end{eqnarray}
as well as the conservation of total atom number, 
\begin{equation}
n=-\frac{1}{\mathcal{S}}\frac{\partial\Omega_{mf}}{\partial\mu},
\end{equation}
where $n=N/\mathcal{S}$ is the number density. At a given temperature,
the ground state has the lowest free energy $F=\Omega_{mf}+\mu N$.

\subsection{Superfluid density tensor}

An important quantity to characterize the anisotropic superfluid properties
of a 2D spin-orbit coupled Fermi gas is the superfluid density tensor.
In the case of BCS pairing, the superfluid density tensor may be analytically
derived within mean-field framework \cite{He2012,Gong2012,Zhou2012},
yet the formalism has not been obtained for a FF superfluid. According
to the definition of the superfluid density, we calculate it by applying
a phase twist to the order parameter, $\Delta_{twist}(\mathbf{v}_{s})=\Delta(\mathbf{r})e^{i\mathbf{q}\cdot\bm{r}}$
, which boosts the system with a uniform superfluid flow at a velocity
$\mathbf{v}_{s}=\hbar\mathbf{q}/2m$ \cite{Zhou2012,Taylor2006,Fukushima2007}.
Here $\Delta(\mathbf{r})$ is the equilibrium FF order parameter.
Physically, only the superfluid component moves under the influence
of the superfluid flow. Thus, as the result of this boost, the thermodynamic
potential assumes the following form in the limit of small velocity,

\begin{equation}
\Omega\left(\mathbf{v}_{s}\right)\simeq\Omega\left(\mathbf{v}_{s}=0\right)+\frac{1}{2}m\mathcal{S}\sum_{ij}n_{s,ij}v_{si}v_{sj},
\end{equation}
where $n_{s,ij}$ ($i,j=x,y$) is the superfluid density tensor. Therefore,
we immediately obtain \cite{Zhou2012,Taylor2006,Fukushima2007},
\begin{equation}
n_{s,ij}=\frac{1}{\mathcal{S}}\frac{4m}{\hbar^{2}}\left[\frac{\partial^{2}\Omega\left(\mathbf{v}_{s}\right)}{\partial q_{i}\partial q_{j}}\right]_{\mathbf{q}=0},\label{eq:ns}
\end{equation}
where $\Omega\left(\mathbf{v}_{s}\right)$ should be calculated with
$\Delta_{twist}(\mathbf{v}_{s})$ in the presence of the phase twist.
The above relation for the superfluid density tensor is rigorous.
In this work, consistent with the mean-field treatment for thermodynamics,
in Eq. (\ref{eq:ns}) we shall approximate the thermodynamic potential
$\Omega\left(\mathbf{v}_{s}\right)$ by its mean-field value $\Omega_{mf}\left(\mathbf{v}_{s}\right)$.

\subsection{The KT-Nelson criterion for $T_{BKT}$}

The BKT transition in 2D is peculiar, associated with the spontaneous
vortex formation. A unique feature of such a transition is a universal
jump in the superfluid density (tensor), characterized by the KT-Nelson
criterion for the critical BKT temperature \cite{Nelson1977}. It
may be explained by using the following simple physical picture for
the spontaneous creation of a single vortex at finite temperature
$T$.

In the absence of spin-orbit coupling and Zeeman fields, let us consider
an \emph{isotropic} Fermi superfluid in a circular disk geometry,
with a radius of $R\rightarrow\infty$. The kinetic energy cost for
creating a single vortex at the origin $\mathbf{r}=\mathbf{0}$ is
simply given by,
\begin{equation}
E_{V}\simeq\frac{1}{2}mn_{s}\int_{\xi}^{R}d^{2}\mathbf{r}\left(\frac{\hbar}{2mr}\right)^{2}=\frac{\hbar^{2}\pi}{4m}n_{s}\ln\left(\frac{R}{\xi}\right),
\end{equation}
where $\xi$ is the size of the vortex core. The associated entropy
can be calculated by the number of distinct positions at which the
vortex can be placed,
\begin{equation}
S_{V}\simeq k_{B}\ln\left(\frac{\pi R^{2}}{\pi\xi^{2}}\right)=2k_{B}\ln\left(\frac{R}{\xi}\right).
\end{equation}
From these two expressions, we see that the free energy associated
with the formation of a single vortex is,
\begin{equation}
F_{V}=E_{V}-TS_{V}\simeq2\left(\frac{\pi}{2}\frac{\hbar^{2}}{4m}n_{s}-k_{B}T\right)\ln\left(\frac{R}{\xi}\right).
\end{equation}
It is clear that the free energy changes its sign at a characteristic
temperature $T_{BKT}$ determined by 
\begin{equation}
k_{B}T_{BKT}=\frac{\pi}{2}\mathcal{J},\label{eq:BKT}
\end{equation}
where $\mathcal{J}=\hbar^{2}n_{s}/(4m)$ is the superfluid phase stiffness.
This is the well-known KT-Nelson criterion \cite{Nelson1977}. As
$\ln(R/\xi)$ diverges in the thermodynamic limit $R\rightarrow\infty$,
the temperature $T_{BKT}$ separates two qualitatively different regimes.
At $T>T_{BKT}$, the free energy is very large and negative, suggesting
the spontaneous creation of a free vortex with either positive or
negative circulation. While at $T<T_{BKT}$, vortices with opposite
circulation will bind together and generate coherence. The spontaneous
creation of free vortex suggests that the loss of the phase coherence
of the system occurs suddenly. It leads to a universal jump in the
superfluid phase stiffness or superfluid density, as can be seen clearly
from the KT-Nelson criterion, Eq. (\ref{eq:BKT}).

In the case of an \emph{anisotropic} superfluid, we need to define
a superfluid density tensor 
\begin{equation}
\mathscr{\mathcal{N}}_{s}=\left[\begin{array}{cc}
n_{s,xx} & n_{s,xy}\\
n_{s,yx} & n_{s,yy}
\end{array}\right].
\end{equation}
The associated superfluid phase stiffness takes the form,
\begin{equation}
\mathcal{J}=\frac{\hbar^{2}}{4m}\left(\det\mathscr{\mathcal{N}}_{s}\right)^{1/2}=\frac{\hbar^{2}}{4m}\sqrt{n_{s,xx}n_{s,yy}},
\end{equation}
where in the last equation, we use the fact that $n_{s,xy}=n_{s,yx}=0$,
which holds for the system considered in this work.

It is worth noting that although Eq. (\ref{eq:BKT}) is obtained by
drawing a simple physical picture, it would be a rigorous criterion
for the BKT transition. Indeed, the KT-Nelson criterion was first
obtained by using a renormalization group analysis \cite{Nelson1977}.
For a microscopic derivation, we may consider the contribution of
the pair fluctuations around the saddle-point solution $\delta\phi\left(\mathbf{q},i\nu_{n}\right)$
to the action $\delta\mathcal{A}$, which, at the \emph{Gaussian}
(quadratic) level, is given by \cite{He2012,He2013,Salasnich2013,Yin2014},

\begin{equation}
\delta\mathcal{A}=\frac{1}{2}\sum_{\mathcal{Q}=\mathbf{q},i\nu_{n}}\left[\delta\phi^{\dagger}\left(\mathcal{Q}\right),\delta\phi\left(-\mathcal{Q}\right)\right]\mathbf{M}\left[\begin{array}{c}
\delta\phi\left(\mathcal{Q}\right)\\
\delta\phi^{\dagger}\left(-\mathcal{Q}\right)
\end{array}\right],
\end{equation}
where the $2\times2$ matrix 
\begin{equation}
\mathbf{M}\equiv\left[\begin{array}{cc}
M_{11}\left(\mathcal{Q}\right), & M_{12}\left(\mathcal{Q}\right)\\
M_{21}\left(\mathcal{Q}\right), & M_{22}\left(\mathcal{Q}\right)
\end{array}\right]
\end{equation}
is the inverse two-particle (pair) propagator and its elements can
be evaluated with the mean-field fermionic Green function $G(\mathbf{k},i\omega_{m})$.
In the case of BCS pairing without the in-plane Zeeman field, the
expression of the inverse pair propagator $\bm{\mathrm{M}}$ can be
analytically obtained \cite{He2013,Salasnich2013}. In particular,
in the limit of long wavelength, the matrix elements of $\bm{\mathrm{M}}$
can be expanded as functions of small $\bm{\mathrm{k}}$ and $\omega$.
By separating the phase fluctuation and amplitude (density) fluctuation,
the low-energy physics of the system can be found to be governed by
the well-known classical spin XY model \cite{He2013,Salasnich2013},
which is the prototype of the BKT physics. In this way, one microscopically
derives the superfluid phase stiffness $\mathcal{J}$ and the KT-Nelson
relation. The resulting expression for the superfluid phase stiffness
\emph{coincides} with the mean-field phase stiffness obtained, for
example, by using the mean-field thermodynamic potential in Eq. (\ref{eq:ns}).
In our FF case, the expression of the superfluid phase stiffness could
be derived in a similar manner. However, in this case, the analytical
expression of the inverse pair propagator $\bm{\mathrm{M}}$ is more
difficult to obtain, although we can numerically sum over the bosonic
Matsubara frequency $i\nu_{n}$. Therefore, to calculate the superfluid
phase stiffness, we prefer to directly use Eq. (17) with a mean-field
thermodynamic potential.

\subsection{Pair fluctuations beyond mean-field}

To close this section, we briefly discuss how to improve the mean-field
theory. An immediate idea is to work out the Gaussian correction to
the action, $\delta\mathcal{A}$, and then use the improved thermodynamic
potential around the saddle point $\Delta(\mathbf{r})=\Delta e^{iQx}$
\cite{Taylor2006,Fukushima2007,Hu2006}, 
\begin{equation}
\Omega_{GPF}=\Omega_{mf}+k_{B}T\sum_{\mathcal{Q}=\mathbf{q},i\nu_{n}}\ln\mathbf{M}\left(\mathcal{Q}\right),
\end{equation}
to calculate the equation of state through the standard thermodynamic
relations and the superfluid density tensor via Eq. (\ref{eq:ns}).
In this way, the thermodynamics and the superfluid density tensor
of the system can be consistently determined at the same level of
approximation. Alternatively, we may also consider using $\Omega_{GPF}$
to determine the chemical potential $\mu$ and then calculate the
superfluid density tensor using the mean-field expression. However,
as the trade-off of this cheap treatment, we may have an inconsistency.
The resulting critical BKT temperature could be less reliable. For
a detailed discussion, we refer to the recent work by Tempere and
Klimin \cite{Tempere2014}.

\section{Results and discussions}

Using the above-mentioned mean-field theoretical framework, we have
systematically explored the low-temperature phase diagram and the
thermodynamic stability of different exotic Fulde-Ferrell superfluid
phases. In our numerical calculations, we take the Fermi wavevector
$k_{F}=\sqrt{2\pi n}$ and the Fermi energy $E_{F}=\hbar^{2}k_{F}^{2}/(2m)$
as the units for wavevector and energy, respectively. For a typical
set of parameters (i.e., default parameters), we use the interaction
parameter $E_{b}=0.2E_{F}$, spin-orbit coupling strength $\lambda=E_{F}/k_{F}$,
in-plane Zeeman field $h_{x}=0.4E_{F}$, out-of-plane Zeeman field
$h_{z}=0.1E_{F}$ and temperature $T=0.05T_{F}$.

\subsection{Low-temperature phase diagrams}

In the recent Letter \cite{Cao2014}, we have discussed the phase
diagram and the appearance of an interesting gapless topological Fulde-Ferrell
superfluid at a weak interaction strength parameterized by $E_{b}=0.2E_{F}.$
Experimentally, it is most likely that the measurement will be carried
out at a stronger interaction strength, where the superfluid transition
temperature is anticipated to be higher. In order to optimize the
experimental condition for observing the gapless topological superfluid,
here we present a systematic study with varying binding energy, from
the weakly interacting BCS side to the strongly interacting BEC-BCS
crossover regime. 

\begin{figure}
\begin{centering}
\includegraphics[clip,width=0.48\textwidth]{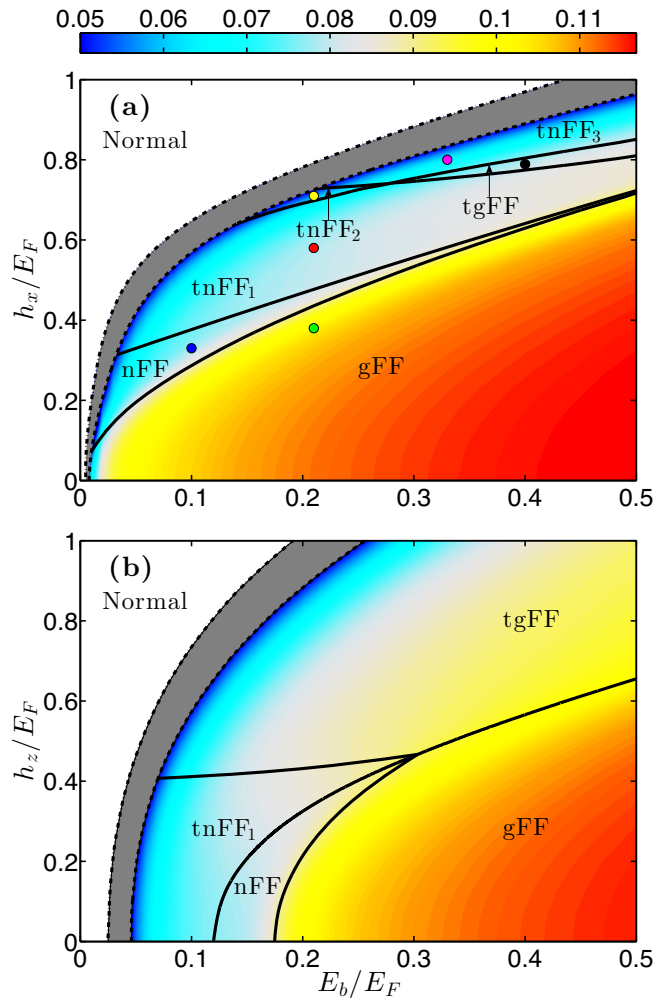} 
\par\end{centering}

\protect\caption{(color online) Phase diagrams of a 2D spin-orbit coupled atomic Fermi
gas at a broad Feshbach resonance and at a typical low temperature
$0.05T_{F}$ with (a) $h_{z}=0.1E_{F}$ or (b) $h_{x}=0.4E_{F}$.
The strength of spin-orbit coupling is $\lambda=E_{F}/k_{F}$. There
are four superfluid phases: gFF, nFF, tnFF and tgFF (whose phase stiffness
$\pi\mathcal{J}/2$ - in units of $E_{F}$ - is illustrated in color),
as well as a pseudogap phase (grey area). We treat the system as a
normal gas (shown in white) when the pairing gap $\Delta<10^{-3}$.
In the gapless topological phase, the notations tnFF$_{1}$, tnFF$_{2}$
and tnFF$_{3}$ distinguish different zero-energy contours in energy
spectrum. For details, see the contour plots in Fig. \ref{fig2}.}

\label{fig1} 
\end{figure}

In Fig. 1, we report two phase diagrams at the typical low temperature
$T=0.05T_{F}$ on the plane of $E_{b}$-$h_{x}$ (a) or $E_{b}$-$h_{z}$
(b). The superfluid phase stiffness $\pi\mathcal{J}/2$ in different
phases is color illustrated and its detailed behavior will be discussed
in the next subsection. The superfluid phases are determined using
the KT-Nelson criterion $\pi\mathcal{J}(T=0.05T_{F})/2>k_{B}T=0.05E_{F}$.
Obviously, there is a pseudogap regime (shown in grey), in which the
pairing order parameter is finite but the superfluid phase stiffness
is not large enough to drive the BKT transition. A better understanding
of the pseudogap phase requires a careful treatment of strong phase
fluctuations. It is out of the scope of the present paper.

\subsubsection{gapless topological transition}

It is known from previous studies \cite{Qu2013,Zhang2013,Cao2014}
that the combined effect of spin-orbit coupling, in-plane and out-of-plane
Zeeman fields may induce several exotic superfluid phases: gapped
FF (gFF), gapless FF (nFF), gapless topological FF (tnFF) and gapped
topological FF (tgFF), classified by considering whether the system
has a bulk-gapped and/or topologically non-trivial energy spectrum.
In the literature, the topological superfluidity was firstly studied
with an out-of-plane Zeeman field only \cite{reviewSOC,Zhang2008}.
In that case, topological phase transition can be driven by increasing
the out-of-plane Zeeman field $h_{z}$ above a threshold 
\begin{equation}
h_{z,c}=\sqrt{\Delta^{2}+\mu^{2}},\label{eq:tpt}
\end{equation}
at which the dispersions of the particle- and hole-branches touch
each other at the single point $\bm{\mathrm{\bm{k}}}=0$, meanwhile
the bulk excitation gap closes. Afterwards, the topology of the Fermi
surface dramatically changes and the excitation gap re-opens \cite{reviewSOC,Hasan2010,Qi2011}.
It is straightforward to understand the single-point closure of the
excitation gap, since the Fermi surface is always rotationally symmetric.
This also implies that the resulting topological superfluid must be
gapped in the bulk. However, such a scenario may be greatly altered
by the presence of a non-zero in-plane Zeeman field, which favors
the FF pairing with a finite center-of-mass momentum and consequently
breaks the rotational symmetry of the Fermi surface. 

In the case of a small in-plane Zeeman field, the rotational symmetry
breaking of the energy spectrum is not significant. Although the system
becomes a FF superfluid, its bulk excitation gap still closes at the
single point $\mathrm{\bm{\mathrm{k}}=0}$, accompanied by the change
of the topology of the Fermi surface. An example is the transition
from gFF to tgFF shown in Fig. \ref{fig1}(b) at large binding energy
$E_{b}>0.3E_{F}$, where the in-plane Zeeman field is effectively
weak. As a result, the picture of the out-of-plane field induced topological
phase transition remains unchanged \cite{Liu2013b,Qu2013,Zhang2013}. 

When the in-plane Zeeman field keeps increasing over a threshold $h_{x,c1}$,
however, the closure of the excitation gap and the change of the topology
of the Fermi surface may not occur at the same time. A gapless superfluid
phase - referred to as nFF - may emerge in the first place at $\bm{\mathrm{\bm{k}}}\neq0$.
The nodal points with $E_{\eta=2}^{\nu}(k_{x},k_{y})=0$ form two
disjoint loops (see, for example, the transition from gFF to nFF in
Fig. \ref{fig1}(a)). The topology of the Fermi surface only changes
when the in-plane Zeeman field further increases up to another critical
value $h_{x,c2}$, at which the two nodal loops connect at $\text{\ensuremath{\bm{\mathrm{k}}}}=0$.
We refer to the previous work Ref. \cite{Cao2014} for a detailed
characterization of the gapless topological transition.

\subsubsection{Binding energy dependence of the phase diagram}

It can now be understood that both the in-plane and out-of-plane fields
can drive the topological phase transition, but the underlying property
of the resulting topological phase, in terms of the gapless or gapped
bulk spectrum, depends critically on the relative strength of the
two fields. The gapless topological FF superfluid (tnFF) intentionally
emerges in the parameter regime where $h_{x}$ is larger enough relative
to $h_{z}$.

This is particularly clear from Fig. \ref{fig1}(a), where we have
fixed the strength of the out-of-plane Zeeman field to $h_{z}=0.1E_{F}$.
The tnFF phase accounts for most of the space for topological phases.
It is remarkable that the window of the tnFF superfluid remains very
significant when the binding energy increases up to $0.5E_{F}$, suggesting
the use of a large interaction strength near Feshbach resonances,
for the purpose of having a larger BKT transition temperature to observe
the exotic tnFF phase. On the contrary, Fig. \ref{fig1}(b) - where
we have fixed the in-plane Zeeman field to $h_{x}=0.4E_{F}$ - clearly
reveals that the gapped topological FF superfluid (tgFF) occupies
most of the space for topological phases, when the out-of-plane Zeeman
field is larger than the in-plane Zeeman field. In this case, the
tnFF phase is restricted to the parameter space with a small out-of-plane
Zeeman field and a weak interaction strength, as one may anticipate.

\subsubsection{Different gapless topological superfluid phases}

\begin{figure}
\begin{centering}
\includegraphics[clip,width=0.48\textwidth]{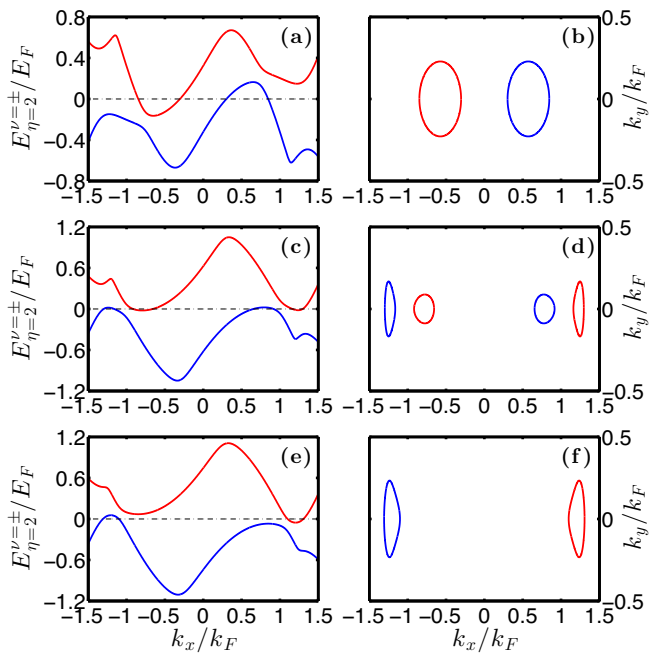} 
\par\end{centering}

\protect\caption{(color online) Dispersion relation of the lower branch $E_{\eta=2}^{\nu}(k_{x},k_{y}=0)$
(left panel, red curves for particle excitations $\nu=+$ and blue
curves for hole excitations $\nu=-$) and the corresponding contour
of zero-energy nodes (right panel). (a) and (b) correspond to the
red point in Fig. \ref{fig1} for the tnFF$_{1}$ phase, (c) and (d)
the yellow point for the tnFF$_{2}$ phase and, (c) and (f) the magenta
point for the tnFF$_{3}$ phase.}

\label{fig2} 
\end{figure}

It is interesting that the gapless topological FF superfluid may be
further classified into different categories (tnFF$_{1}$, tnFF$_{2}$
and tnFF$_{3}$), according to the number and position of its disjoint
loops of nodal points, as shown in the right panel of Fig. \ref{fig2}.
The tnFF$_{1}$ superfluid is most common and has two nodal loops,
one for the particle branch (red loop) and another for the hole branch
(blue loop). The tnFF$_{3}$ superfluid also has two nodal loops.
However, the loops for the particle and hole branches exchange their
position. It occurs only at large in-plane Zeeman field and binding
energy. The tnFF$_{2}$ seems to connect the tnFF$_{1}$ and tnFF$_{3}$
phases. It has four disjoint nodal loops and exists only in a very
narrow parameter space (see, for example, Fig. \ref{fig1}(a)). We
note that the two gapless topological phases, tnFF$_{1}$ and tnFF$_{3}$,
may also be intervened by a \emph{gapped} topological phase, in which
there is no nodal loop at all.

\subsection{Superfluid density}

Having determined the low-temperature phase diagram, we are in position
to understand the superfluid density and the critical BKT temperature
of different superfluid phases, which have been only briefly mentioned
in our previous Letter \cite{Cao2014}. In the presence of spin-orbit
coupling, it is known that the superfluid density is a tensor \cite{He2012,He2013}.
We then have to consider both diagonal elements of the superfluid
density tensor, $n_{s,xx}$ and $n_{s,yy}$.

\begin{figure}
\begin{centering}
\includegraphics[clip,width=0.48\textwidth]{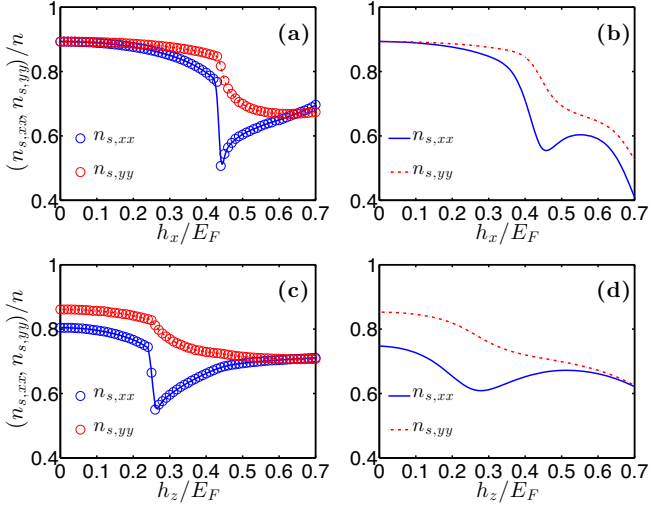} 
\par\end{centering}

\protect\caption{(color online) Diagonal elements of the superfluid density tensor
as a function of $h_{x}$ and $h_{z}$ at zero temperature (left panel)
and at a finite temperature $T=0.05T_{F}$ (right panel). The superfluid
density is measured in units of the total density $n=k_{F}^{2}/(2\pi)$.
In (a) and (b), the out-of-plane Zeeman field strength $h_{z}=0.1E_{F}$.
In (c) and (d), the in-plane Zeeman field strength $h_{x}=0.4E_{F}$.
Other parameters are $E_{b}=0.2E_{F}$ and $\lambda=E_{F}/k_{F}$.}

\label{fig3} 
\end{figure}

In Fig. \ref{fig3}, we present the Zeeman field dependence of $n_{s,xx}$
and $n_{s,yy}$ at zero temperature (left panel, a and c) and at a
finite temperature $T=0.05T_{F}$ (right panel, b and d). In general,
as a consequence of the in-plane Zeeman field applied along the $x$-axis,
$n_{s,xx}$ is smaller than $n_{s,yy}$, except at extremely low temperature
and sufficiently large Zeeman fields. 

At zero temperature, $n_{s,xx}$ initially decreases with increasing
Zeeman fields and exhibits a sudden drop when the system evolves from
the gFF phase into the nFF phase at the threshold $h_{x,c1}$ (or
$h_{z,c1}$). At $h_{x}>h_{x,c1}$ (or $h_{z}>h_{z,c1}$) it then
rises up gradually and is always enhanced by the Zeeman field. Apart
from the discontinuous jump, similar Zeeman-field dependence of the
superfluid density has been reported for a gapped BCS topological
superfluid across the topological phase transition \cite{He2013}.
Compared with the non-monotonic field dependence of $n_{s,xx}$, we
always find that $n_{s,yy}$ decreases continuously with increasing
the Zeeman field. Instead of the sudden drop, a kink is observed at
the transition from the gFF phase to the nFF phase. 

The behavior of the superfluid density is profoundly affected by a
nonzero temperature. Already at $T=0.05T_{F}$, the discontinuous
drop in $n_{s,xx}$ is smoothed out, leaving a broad minimum with
a width $\Delta h_{x,z}\sim2k_{B}T=0.1E_{F}$. Moreover, at the large
Zeeman field $h_{x,z}\sim0.6E_{F}$, $n_{s,xx}$ starts to decrease
with increasing the Zeeman field. At even higher temperature (not
shown in the figure), the local minimum in $n_{s,xx}$ may disappear.

\begin{figure}
\begin{centering}
\includegraphics[clip,width=0.48\textwidth]{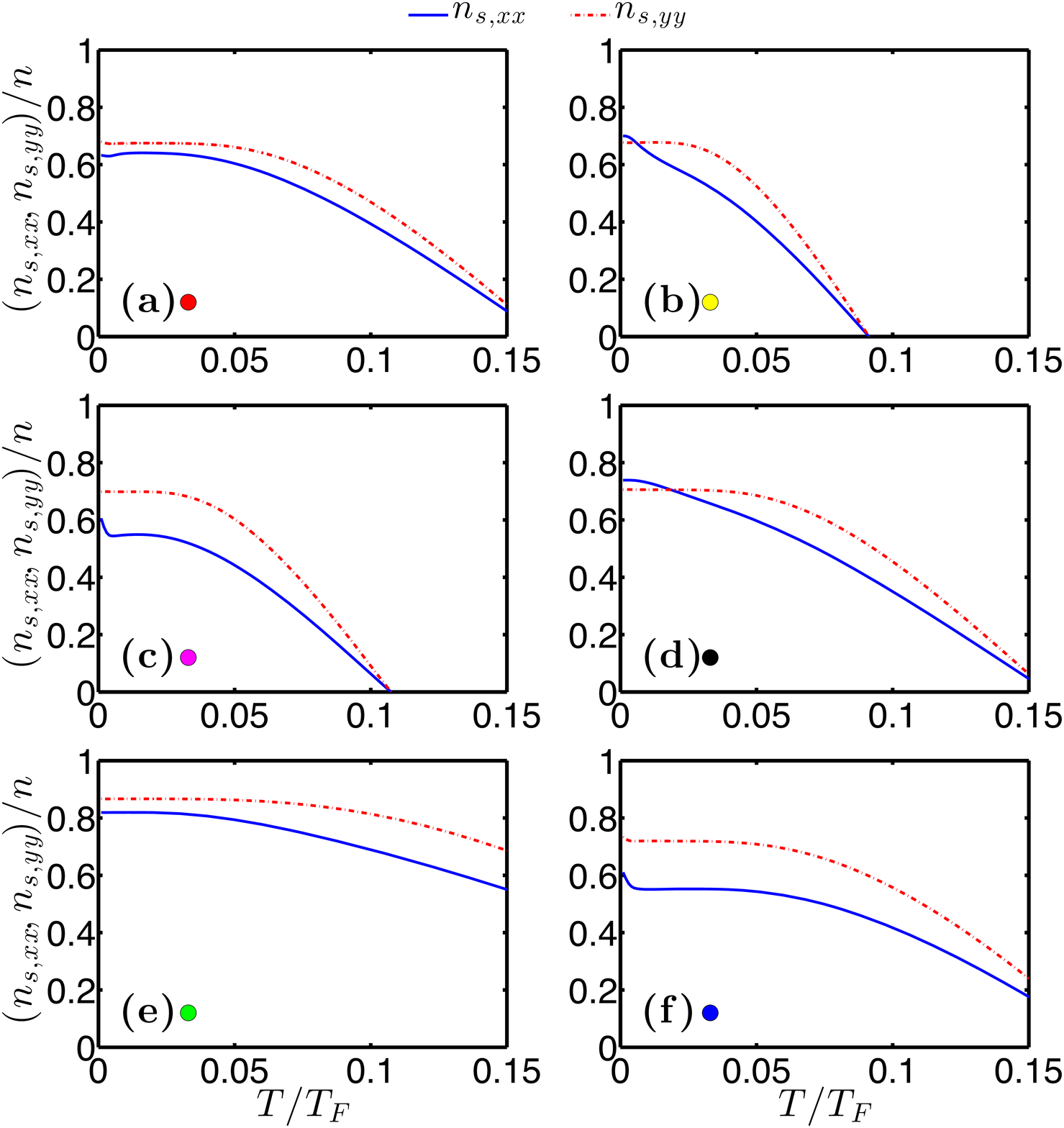} 
\par\end{centering}

\protect\caption{(color online) Temperature dependence of the diagonal elements of
the superfluid density tensor, at the six points shown in Fig. 1(a):
(a) $E_{b}=0.21E_{F}$ and $h_{x}=0.58E_{F}$, the tnFF$_{1}$ phase;
(b) $E_{b}=0.21E_{F}$ and $h_{x}=0.71E_{F}$, the tnFF$_{2}$ phase;
(c) $E_{b}=0.33E_{F}$ and $h_{x}=0.8E_{F}$, the tnFF$_{3}$ phase;
(d) $E_{b}=0.4E_{F}$ and $h_{x}=0.789E_{F}$, the tgFF phase; (e)
$E_{b}=0.21E_{F}$ and $h_{x}=0.2E_{F}$, the gFF phase; and (f) $E_{b}=0.1E_{F}$
and $h_{x}=0.33E_{F}$, the nFF phase. The superfluid density is measured
in units of the total density $n=k_{F}^{2}/(2\pi)$. Other parameters
are $h_{z}=0.1E_{F}$ and $\lambda=E_{F}/k_{F}$.}

\label{fig4} 
\end{figure}

In Fig. \ref{fig4}, we report the temperature dependence of the superfluid
density at six typical sets of parameters, which correspond to different
superfluid phases at $T=0.05T_{F}$, as shown in Fig. 1(a). $n_{s,xx}$
and $n_{s,yy}$ decrease as temperature increases, in agreement with
the common idea that the superfluid component should be gradually
destroyed by thermal excitations. It is remarkable that for the gapless
topological tnFF$_{1}$ phase (see Fig. \ref{fig4}(a)), the superfluid
density does not decrease rapidly with increasing temperature, implying
a sizable critical BKT transition temperature for its experimental
observation, as we shall discuss in greater detail in the next subsection.
In contrast, the superfluid density of other two gapless topological
phases (tnFF$_{2}$ and tnFF$_{3}$ in Figs. \ref{fig4}(b) and \ref{fig4}(c),
respectively) is more sensitive to temperature and vanishes at $T\sim0.1T_{F}$,
probably due to their large Zeeman fields.

\subsection{Critical BKT temperature and finite-temperature phase diagrams}

\begin{figure}
\begin{centering}
\includegraphics[clip,width=0.48\textwidth]{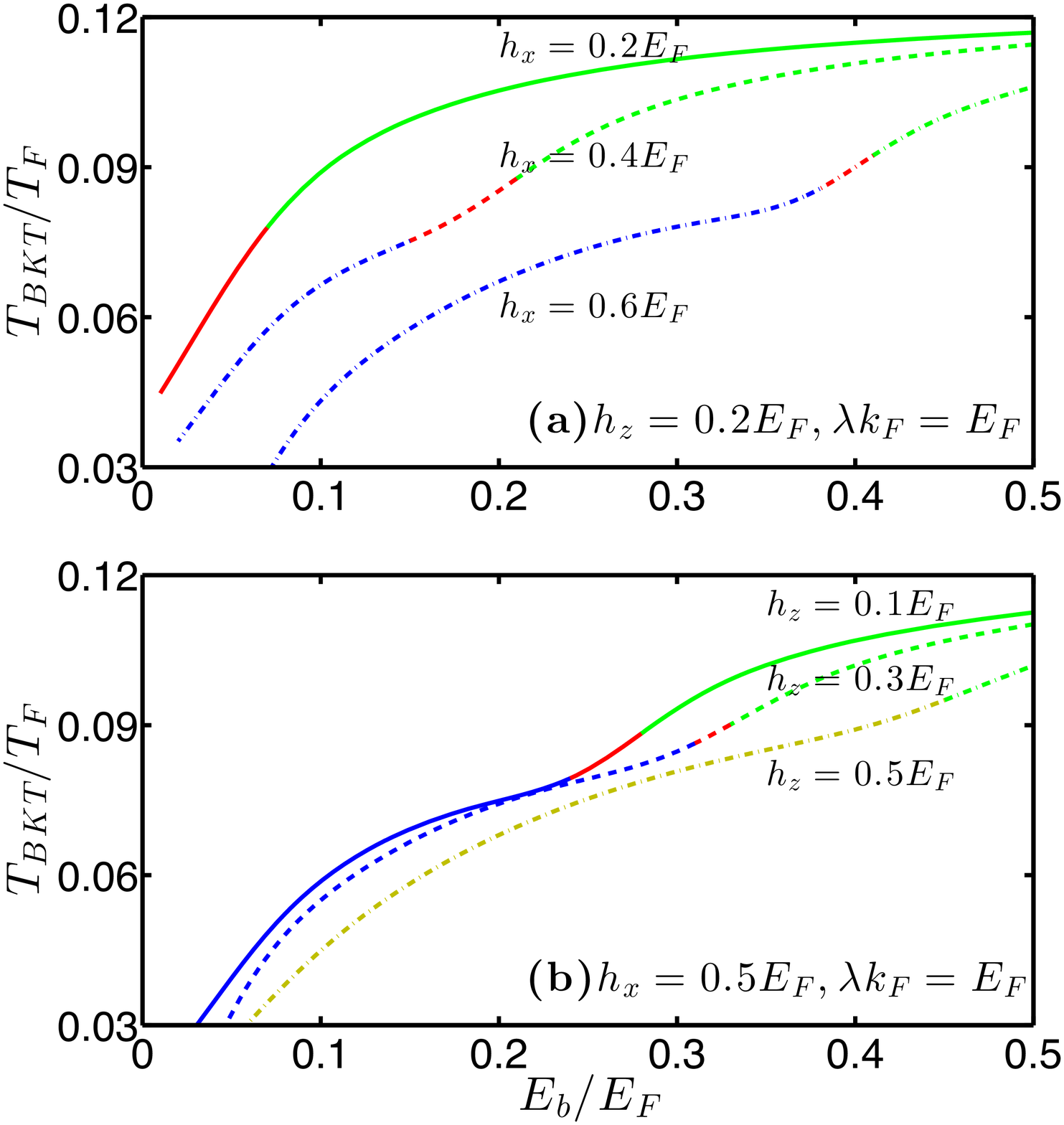} 
\par\end{centering}

\protect\caption{(color online) The critical BKT transition temperature as a function
of the binding energy $E_{b}$ at different in-plane Zeeman fields
(a) or out-of-plane Zeeman fields (b). Here and in the next two figures,
the color green, red, blue and yellow in the curves denote the superfluid
phase gFF, nFF, tnFF and tgFF, respectively.}

\label{fig5}
\end{figure}
\begin{figure}
\begin{centering}
\includegraphics[clip,width=0.48\textwidth]{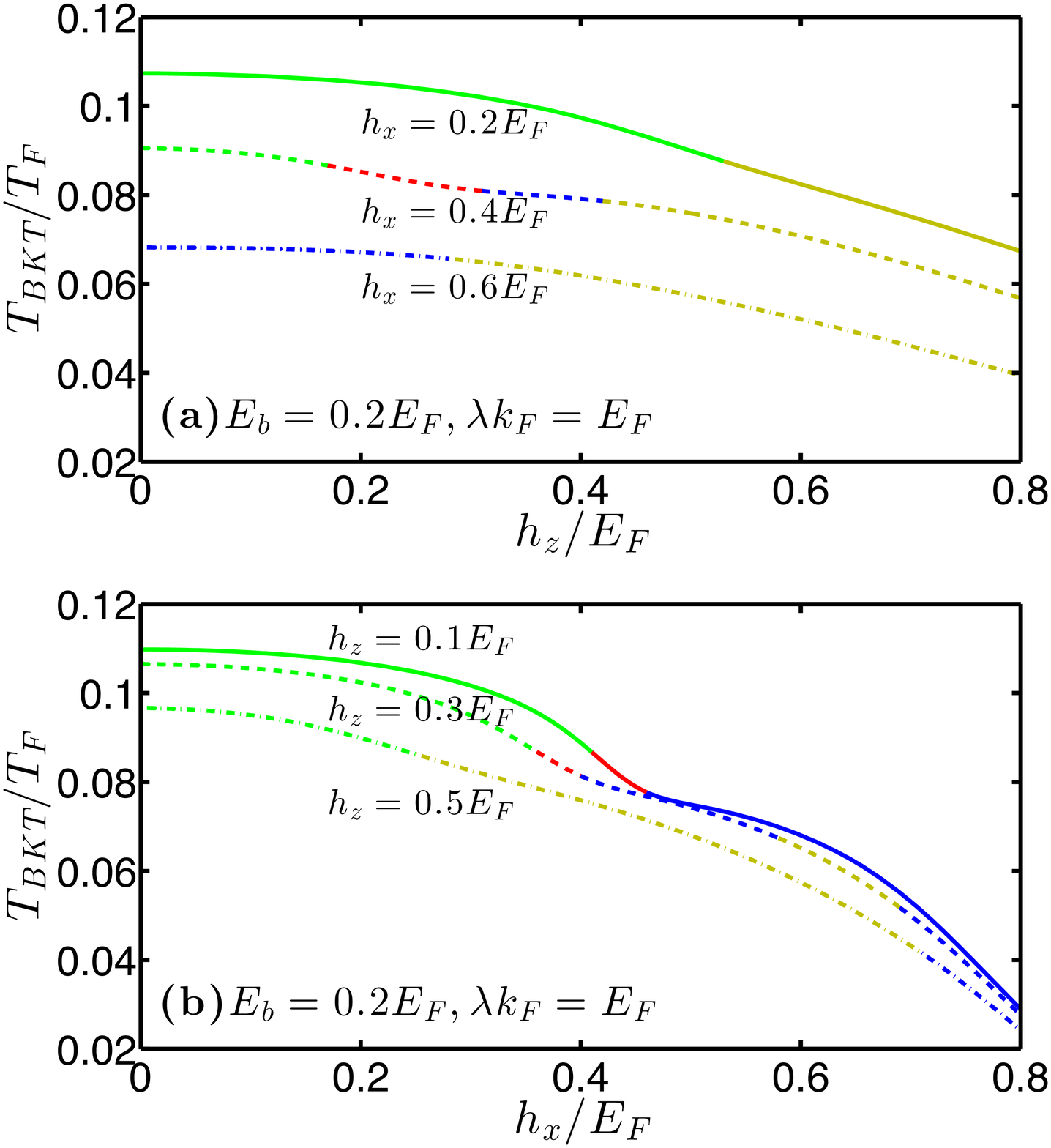} 
\par\end{centering}

\protect\caption{(color online) The critical BKT transition temperature as a function
of the out-of-plane Zeeman field $h_{z}$ (a) or the in-plane Zeeman
field $h_{x}$ (b).}

\label{fig6}
\end{figure}
\begin{figure}
\begin{centering}
\includegraphics[clip,width=0.48\textwidth]{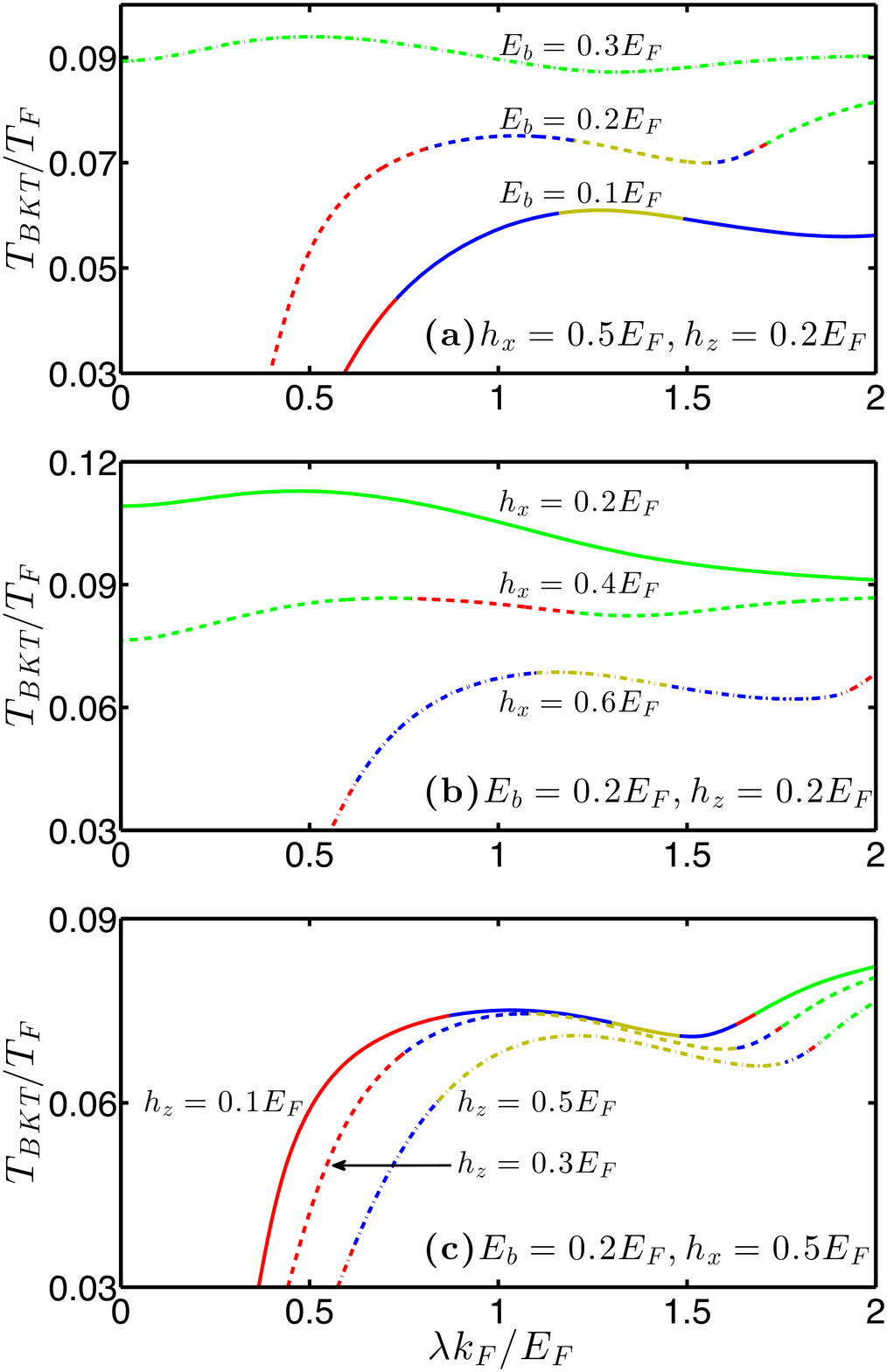} 
\par\end{centering}

\protect\caption{(color online) The critical BKT transition temperature as a function
of the spin-orbit coupling strength $\lambda$ at different binding
energies (a), in-plane Zeeman fields (b) and out-of-plane Zeeman fields
(c). }

\label{fig7} 
\end{figure}

We now turn to consider the critical BKT temperature, which is determined
by the KT-Nelson criterion, 
\begin{equation}
k_{B}T_{BKT}=\frac{\pi\hbar^{2}}{8m}\left[n_{s,xx}\left(T_{BKT}\right)n_{s,yy}\left(T_{BKT}\right)\right]^{1/2}.\label{eq:BKT2}
\end{equation}
In the above equation, we have explicitly written down the temperature
dependence of the superfluid density, in order to emphasize the fact
that the critical BKT temperature should be solved self-consistently.
In Figs. \ref{fig5}, \ref{fig6} and \ref{fig7}, we show the results
as a function of the binding energy, Zeeman fields and spin-orbit
coupling strength, respectively. These results should be regarded
as finite-temperature phase diagrams, as they show clearly which kind
of superfluid phases is preferable when temperature decreases. In
the curves, we use different colors to distinguish different \emph{emerging}
superfluid phases: green for the gFF phase, red for the nFF phase,
blue for the tnFF phase and finally yellow for the tgFF phase. It
is clear that all the four FF superfluid phases have significant critical
BKT temperature except for the parameter regime with very small binding
energy $E_{b}$ and/or spin-orbit coupling strength $\lambda$, or
with very large in-plane Zeeman field $h_{x}$ and/or out-of-plane
Zeeman field $h_{z}$.

As illustrated in Fig. \ref{fig5}, the critical BKT temperature $T_{BKT}$
always increases monotonically with increasing the binding energy
$E_{b}$, as the pairing and superfluidity are enhanced at strong
interatomic interactions. The binding energy is the dominant factor
in forming Cooper pairs. With a small binding energy, the system is
mainly of fermionic character. On the contrary, with a sufficiently
large binding energy, the system tends to act as a gas of bosons.
Therefore, with increasing the binding energy up to some points, the
system would lose its fermionic character near the BEC-BCS crossover
(i.e., $E_{b}\sim0.5E_{F}$) and hence should become topologically
trivial. Indeed, at large binding energy we observe that the system
always approaches the topologically trivial gFF phase. The topological
phase, either gapless (tnFF in blue) or gapped (tgFF in yellow), is
favored at small binding energy, where the critical BKT temperature
is lower. Nevertheless, we find that by suitably tuning the parameters,
it is possible to have a gapless topological tnFF phase with a sizable
critical BKT temperature $T_{BKT}\sim0.09T_{F}$ for the binding energy
up to $E_{b}\simeq0.4E_{F}$ (see, for example, the dot-dashed line
at the bottom of Fig. \ref{fig5}(a)). This temperature is clearly
within the reach in current cold-atom experiments \cite{Ries2014}.

On the other hand, the critical BKT temperature decreases monotonically
with increasing the Zeeman field, either in-plane or out-of-plane,
as shown in Fig. \ref{fig6}. It is readily seen that with decreasing
temperature the system would first turn into either the tnFF or tgFF
phase at sufficiently large in-plane Zeeman field $h_{x}$ or out-of-plane
field $h_{z}$, respectively. While at low Zeeman fields, the topologically
trivial gFF phase is preferable. This agrees the observation we made
in discussing the low-temperature phase diagrams in Fig. \ref{fig1}.

It is worth noting that, one may use the binding energy dependence
or the Zeeman field dependence of the critical BKT temperature to
identify different emerging superfluid phases. This is particularly
clear for the gapless tnFF and nFF phases, as the curvature of the
$T_{BKT}$ curve for those phases behaves quite differently. For the
tnFF phase, the curve is concave; while for the nFF phase, it is convex.
This change in curvature (i.e, from concave to convex) seems to be
related to the local minimum in the superfluid density component $n_{s,xx}$
that we have reported earlier in Fig. \ref{fig3}.

We now discuss the critical BKT temperature as a function of the spin-orbit
coupling strength $\lambda$, as shown in Fig. \ref{fig7}. Compared
with the binding energy dependence and Zeeman field dependence, the
dependence of $T_{BKT}$ on the spin-orbit coupling strength is non-monotonic
and the emerging superfluid phases can re-appear with increasing the
coupling strength. Therefore, the $T_{BKT}$ curve is more subtle
to understand. Nevertheless, we may identify that the topologically
trivial gFF superfluid phase tends to be favorable at large spin-orbit
coupling. This is because the pairing gap is usually enhanced by the
spin-orbit coupling, which makes the topological phase transition
much more difficult to occur (cf. Eq. (\ref{eq:tpt})). At small spin-orbit
coupling, on the other hand, the critical BKT temperature may dramatically
decrease to zero, particularly at a small binding energy and/or a
large Zeeman field. Thus, for the purpose of observing the gapless
topological tnFF phase, experimentally it seems better to use an intermediate
spin-orbit coupling strength, i.e., $\lambda\sim E_{F}/k_{F}$.

\section{Conclusions}

In summary, we have presented a systematic investigation of the Berezinskii-Kosterlitz-Thouless
transition in a spin-orbit coupled atomic Fulde-Ferrell superfluid
in two dimensions. We have calculated the superfluid density and superfluid
transition temperature of various Fulde-Ferrell superfluids. We have
paid special attention to an interesting gapless topological Fulde-Ferrell
superfluid and have clarified that, by suitably tuning the external
parameters - for example, the interatomic interaction strength, in-plane
and out-of-plane Zeeman fields, and spin-orbit coupling strength -
its observation is within the reach in current cold-atom experimental
setups.

Our investigation is based on the mean-field theoretical framework,
which is supposed to be applicable to a weakly interacting two-dimensional
Fermi gas (i.e., the binding energy $E_{b}\leq0.2E_{F}$). For a more
reliable and quantitative description, in future studies it would
be useful to take into account the strong phase fluctuations by using
many-body \textit{T}-matrix theories \cite{Hu2006,Hu2008,Hu2010}. 
\begin{acknowledgments}
X.J.L. and H.H. were supported by the Australian Research Council
(ARC) (Grants Nos. FT140100003, FT130100815, DP140103231 and DP140100637)
and the National Key Basic Research Special Foundation of China (NKBRSFC-China)
(Grant No. 2011CB921502). L.H. was supported by US Department of Energy
Nuclear Physics Office. G.L.L. was supported by the National Natural
Science Foundation of China (NSFC-China) (Grant Nos. 11175094, 91221205)
and the NKBRSFC-China (Grant No. 2011CB921602). 

\textit{Note added}. Recently, a similar publication by Xu and Zhang
became public \cite{Xu2014}. Results qualitatively agree where applicable.\end{acknowledgments}

\end{document}